

\magnification=1200
\hoffset=0.0 true cm

\voffset=0.0 true cm
\vsize=21.0 true cm
\hsize=16.5 true cm

\baselineskip=10pt     
\parskip=5pt           

\parindent=22pt
\raggedbottom

\def\pp{\noindent\parshape 2 0.0 truecm 17.0 truecm 0.5 truecm 16.5 truecm}

\font\bigbf=cmb10 scaled \magstep2

\def\x{X-ray~}
\def\lr{XR~}
\def\lxr{nXR~}

\def\etal{{\it et~al.~}}
\def\lsim{\hbox{ \rlap{\raise 0.425ex\hbox{$<$}}\lower 0.65ex\hbox{$\sim$} }}
\def\gsim{\hbox{ \rlap{\raise 0.425ex\hbox{$>$}}\lower 0.65ex\hbox{$\sim$} }}
\def\pn{\par\noindent}


\null\vskip 1.0 truecm

\centerline{\bigbf LOW MASS X-RAY BINARIES IN GLOBULAR CLUSTERS:}
\bigskip
\centerline{\bigbf A NEW METALLICITY EFFECT}
\bigskip
\centerline{\bigbf }
\bigskip
\bigskip

\centerline {\bf M. Bellazzini$^1$, A. Pasquali$^2$, L. Federici$^3$}
\bigskip
\centerline{\bf F.R. Ferraro,$^3$ F. Fusi Pecci $^3$}

\bigskip
\bigskip
\centerline{$1:$ e-mail: 37929::bellazzini; bellazzini@astbo3.bo.astro.it}
\centerline{$2:$ e-mail: pasquali@arcetri.astro.it}
\centerline{$3:$ e-mail: 37929::luciana; luciana@astbo3.bo.astro.it}
\centerline{$4:$ e-mail: 37929::ferraro; ferraro@astbo3.bo.astro.it}
\centerline{$5:$ e-mail: 37929::flavio; flavio@astbo3.bo.astro.it}
\centerline{fax: 39-51-259407}

\bigskip
\bigskip

\centerline{$^1$ {\it Dipartimento di Astronomia, Via Zamboni 33,
I-40126 Bologna, Italy. }}
\par\noindent
\centerline{$^2$ {\it Dipartimento di Astronomia e Scienza dello Spazio,
Largo E. Fermi 5, I-35100 Firenze, Italy}}
\par\noindent
\centerline{$^3$ {\it Osservatorio Astronomico di Bologna, Via Zamboni 33,
I-40126 Bologna, Italy.}}

\bigskip
\bigskip
\bigskip

\centerline{The Astrophysical Journal {\it (in press)}}
\bigskip
\centerline{Proofs to: Flavio Fusi Pecci}
\bigskip

\vfill\eject

\baselineskip=20pt

\centerline{\bf ABSTRACT}
\bigskip
\noindent {
Galactic Globular Clusters (GCs) containing bright \x sources ($L_x  > 10^{36}
erg/s$), commonly associated with Low Mass \x Binaries (LMXBs),
are found to be significantly denser and more metal-rich than
normal non-X-ray clusters both in the Galaxy and in M31.

Within a framework where LMXBs in GCs are generated via tidal captures
in high-density clusters and $(2+1)$ encounters in low-density globulars,
the higher incidence of LMXBs with increasing metallicity
is shown to be {\it intrinsic} and not just a by-product of other effects.
\pn
Two possible mechanisms are examined: the first one assumes a dependence of
the cluster IMF on metallicity as recently
published in the literature. The observed LMBXs, more frequently occurring in
metal rich clusters, agrees with the predicted number of NS only if metallicity
accounts for a minor contribution to the observed variation of the IMF slope.
Other alternatives explored, such as the total variation of the
observed IMF slopes is due to (a) just metallicity and (b) the combination of
metallicity and position in the Galaxy lead to a clear-cut disagreement
with the data.
In turn, this result may indicate a flatter dependence of the cluster's IMF on
metallicity than that deduced from observed cluster luminosity functions.

The second mechanism assumes that, at fixed cluster density, the rate of tidal
captures depends on radius and mass of the capturing star.
Based on standard stellar models, stars with higher metal content have
wider radii and higher masses, hence the rate of tidal captures increases
with increasing metallicity. Moreover, since for fixed
binary separation and masses of the two components, metal-rich stars fill
more easily the Roche lobe, as their radii are larger, there is an
additional ``evolutionary'' reason to favour a higher incidence of LMXBs
in metal-rich clusters. From the order of magnitude computations made,
this new effect
by itself could explain the observed ratio of 4 between the frequencies of
X-ray clusters in the metal-rich and metal-poor groups we observationally
determined. However, there is no reason to exclude that both mechanisms
can be at work.
}

\bigskip\noindent
{\it Subject index headings:}
Clusters: Globular (126); Stars: Binaries; X-ray sources;

\vfill\eject
\noindent
\centerline {\bf 1. INTRODUCTION}

\bigskip\noindent
Bright \x sources ($L_x > 10^{36}$ erg/s) are thought to be
binary systems where a Neutron Star is accreting mass from a
companion (Lightman and Grindlay, 1982), and they are commonly
known as Low Mass \x Binaries (LMXBs).

In a recent review, Grindlay (1993) lists 12 Galactic Globular Clusters (GGCs)
from the {\it Einstein} and {\it ROSAT} surveys which contain LMXBs.
The LMXB globulars belong mostly to the galactic
disk (9/12), and exhibit, with the exception of M15, a quite high metal
content and, with the exception of NGC 6712, high central density.
Surprisingly enough, the above sample is not significantly overabundant in
post-core collapse (PCC) clusters with respect to the whole GGC
population.
In fact, the fraction of LMXB globulars which seem to have undergone core
collapse is the same as in the whole Galactic cluster system (Grindlay 1993).
If cluster central density were the only crucial
parameter favouring the formation of \x binaries, one would
expect a very tight correlation between PCC morphology and LMXB
population. Since this is not the case,
another parameter may influence, besides high stellar density, the production
of \x binaries in globulars.

In this note a comparison between the ``X-ray'' (\lr) and
``non-X-ray'' (\lxr) cluster populations in the Galaxy and in M31 is first
presented, and then
a working hypothesis which suggests that the ``second parameter''
involved in the production of LMXBs is actually the cluster metallicity is
proposed.

\bigskip\noindent
\centerline {\bf 2. THE DATA-BASE}

\bigskip\noindent
{\bf 2.1.  The Galaxy}

\smallskip\noindent
The \x survey of the Galaxy secured by {\it Einstein} and {\it ROSAT}
satellites
can be considered exhaustive at the luminosity level
$L_x > 10^{36}$  erg/s. Hence,  the 12 GGCs  listed in his Table 1 by Grindlay
(1993) are adopted as {\it bona fide} LMBX globulars.

To carry out useful comparisons, a sample of 122
\lxr GGCs was then taken from Zinn (1985). For all clusters,
metallicities are taken from Zinn (1985), Armandroff and Zinn (1988),
Armandroff (1989) and  Armandroff \etal (1992).
Central luminosity densities, distances from the center of the Galaxy and from
the Galactic Plane are from Djorgovski (1993), while central velocity
dispersions and M/L ratios are from Pryor and Meylan (1993).

\medskip\noindent
{\bf 2.2. M31}

\smallskip\noindent
M31 is less obvious to deal with and an accurate selection
of the samples must be done before making any further analysis and comparison.

The \x surveys carried out with {\it Einstein} in M31 and presented
and discussed by Long and van Speybroeck (1983, hereafter--LVS), Crampton \etal
(1984--CCHSV), Trinchieri and Fabbiano (1991--TF) and those made with
{\it ROSAT}
by Primini, Forman and Jones (1993--PFJ) covered at sufficiently
high resolution only quite small areas centered on the bulge of the galaxy.
As a consequence, the available samples are surely incomplete.

In order to be homogeneous, the present selection was limited to sources found
with the HRI-cameras operated on  both satellites, wich were sufficiently
similar in spatial resolution and wavelength range. In their analysis,
TF consider a sort of strip oriented along the M31 major axis, extended
$\sim 2$ deg in DEC and $\sim 3$ min in RA (see their Fig. 1).
Within this region, we selected a sample of 26 \lr globulars,
most of which have multiple identifications made by various authors
(LVS, CCHSV, TF, PFJ). These identifications were checked by independently
cross-correlating all the detected \x sources listed by the different
authors with the revised catalog of M31 globular cluster candidates
(hereafter {\it Bo-catalog}) constantly updated by the Bologna Group
(Battistini \etal 1980, 1987, 1993, Federici \etal 1993,
Fusi Pecci \etal 1993b). Adopting error boxes either slightly smaller or
larger than previously used by other authors, exactly the same 26
identifications were found. It is therefore safe to assume
this sample is reliable enough to be used as truly representative
of the \lr globulars in M31. Table 1 reports the list of the 26 \lr GCs
included in our sample and some useful data discussed
below.

In order to construct a corresponding sample of \lxr clusters
to make comparisons and tests as for the Milky Way population,
the list of cluster candidates (classes A,B) identified in the same region
considered in the \x survey was also extracted from the {\it Bo-catalog}.
In total,
excluding all candidates whose spurious nature has so far been
verified, 171 objects were found.
The much larger distance which separates us from M31 than from our own Galaxy
GCs makes to measure intrinsic structural parameters and metallicity a much
harder task.
In fact, at M31 ($d\sim700 Kpc$), 1 $pc$ corresponds
to about 0.3 $arcsec$, and only HST-data can yield fully reliable structural
parameters for the M31 clusters (Fusi Pecci \etal 1993a).

However, in order to be able to make educated guesses, a rough
estimate of the intrinsic structure of the M31 clusters was made following
the same procedures previously adopted by our group
(Buonanno \etal 1982, Battistini \etal 1982).
The chosen observable is simply the half-width at $1/4$ of the height
--~$W_{1/4}$~-- of the two-dimensional fit of the cluster image, normalized
to the arbitrary scale of a reference plate (see also Battistini \etal
1987, Battistini \etal 1994). Using the available data, this quantity
is now available for almost all the \lr (25 out of 26) and
\lxr (158 out of 171) clusters. We report in Table 1
(column 4) the individual values for \lr globulars.

The whole data set for the remaining M31
clusters, and for the whole Galaxy sample, can be found in the above quoted
papers or can be obtained upon request and therefore is not reproduced here.
Note that here ~$W_{1/4}$~ runs opposite to the density,
{\it i.e.} clusters having small ~$W_{1/4}-$values are
expected to be actually highly concentrated objects.
The substantial reliability of these estimates, at least at a first order,
has been repeatedly checked with other measures from the ground
(Crampton \etal 1985, Cohen and Freeman 1991) and with HST
(Fusi Pecci \etal 1993a). On the other hand, nothing better is
available at this stage for the considered sample.

Metallicity estimates
([Fe/H]) by Huchra, Brodie and Kent (1991) are based on accurate spectral
index calibrations and can be assumed
to be free from any bias induced by the quite strong reddening affecting
many of the considered objects and were therefore adopted.
However, we must note that while Huchra, Brodie and Kent (1991) list
metallicities for 21 out of
26 \lr globulars ({\it i.e.} $\sim 80\%$ of the total sample)
, they give values for only 68 \lxr clusters ($\sim40\%$ of the total
sample). Other estimates
based on proper calibrations of photometric colors have been made
by various authors including our group, but since reddening effects
may have influenced the observables, they were not used in order to prevent
contamination of the sample.\noindent

Finally, the projected galactocentric distances (R$_{GC}$ \- in Kpc)
are taken from Battistini \etal 1987 ($d_{M31}$ = 651 Kpc).

\bigskip\noindent
\centerline {\bf 3. RESULTS}

\bigskip\noindent
Having at hand the data for the four sets of clusters described
above, one can immediately compute average values of the various
parameters and verify whether any significant difference emerges between the
various groups.
In Table 2, therefore, the mean values obtained for each considered
parameter for \lr and \lxr globulars in the Galaxy and in M31,
respectively, are listed. For the metallicities,  the weighted means
and the corresponding associated errors were computed making use of the
available estimates of the uncertainties for each single object.
\medskip\noindent
{\bf 3.1. Cluster Density}

\smallskip\noindent
The difference in
average density between the two groups in the Galaxy ($<Log \rho_0>_{\lxr}
= 3.41 \pm 0.16$ and $<Log \rho_0>_{\lr} = 4.93 \pm 0.26$) it is quite large.
A Kolmogorov-Smirnov (KS) test shows that the two samples are drawn from
different populations at a $\sim 99.7\%$ ($\sim 3\sigma$) confidence level.

Keeping in mind that ~$W_{1/4}$~ is anticorrelated to the
cluster density, in M31 ~$W_{1/4}^{\lr} = 4.36\pm0.07$~ has to be compared
with ~$W_{1/4}^{\lxr} = 5.00\pm0.04$~.
The KS-test applied to the two distributions shows that they are different
at $\sim 98.8\%$ ({\it i.e.} $2.5 \sigma$) confidence level.
Hence, though less significant, but it is not a surprise given the
lower quality of the input data, it seems confirmed that also the M31
globulars containing LMXBs are denser than normal clusters.

In Figure 1$a,b$ the cumulative distributions for the two
considered quantities ($Log \rho_0$ in the Galaxy and ~$W_{1/4}$ in
M31) have been plotted, where the {\it solid line} is for the
\lr globulars and the {\it dotted line} for the \lxr clusters, respectively.

\medskip\noindent
{\bf 3.2. Distance from the Galactic Center and the Galactic Plane}

\smallskip\noindent
As shown also from the cumulative distribution in Figure 2$a$,
the KS-test applied to the distribution of the distances from the Galactic
Center ($R_{GC}$) for \lr and \lxr clusters indicates, at $\sim 92\%$
significance level, that they are drawn from different parent populations.
This marginal indication could also be simply a by-product of
the fact that high density clusters are more frequent at small $R_{GC}$.
\par
Concerning the height with respect to the Galactic Plane
($|Z_{GP}|$, see Figure 3), the probability that the two samples
are extracted from the same parent population is only $\sim 3\%$. However,
one should note that the detected difference is mostly based
on a few \lr clusters located very close to the Galactic Plane
(See Sect. 4.2. for further discussion).
\par
For the two groups (\lr and \lxr) in M31 the situation appears similar.
In fact, the KS-test applied to the distributions of
Figure 2$b$, the two populations are different at a
$\sim 99.1\%$ confidence level ({\it i.e.} $2.6 \sigma$).
However, this effect could also be due to a possible bias
in the cluster selections and in the mapping of the \x observations.
For instance, while in the Galaxy both the optical and \x cluster
surveys are probably complete, many low luminosity clusters are
surely still missing in the inner regions of M 31, and this would alter the
relative distributions (see Battistini \etal 1980, 1987, Federici \etal 1993,
Fusi Pecci \etal 1993b for discussions on selection bias in the M31 cluster
survey). Unfortunately, the available data make it impossible for M31
to carry out any test concerning $|Z_{GP}|$, as it was done for the Galaxy.

\medskip\noindent
{\bf 3.3. Metallicity}

\smallskip\noindent
The weigh\-ted mean me\-tal\-licity for the 12 \lr GCs in the Ga\-laxy
turns out to be\pn
$<[Fe/H]>_{\lr}= -1.17 \pm 0.04$. This figure is higher than that
obtained for
the \lxr sample $<[Fe/H]>_{\lxr} = -1.41 \pm 0.01$, which is in good agreement
with the average value determined for the whole cluster system in the
Galaxy for instance by using data from Zinn (1985) or Huchra \etal (1991).
A KS-test applied to the two distributions indicates
that the hypothesis that they are extracted from the same population can
be rejected at $99.9\%$ confidence level ({\it i.e.} $> 3\sigma$). Thus
the evidence is very strong and, as said, not new.

Since LMXB clusters are mostly disk objects, and are usually metal-rich,
one might interpret the detected metallicity difference to be {\it not
germane}, but simply a secondary consequence of the primary dependence on
density and membership to the disk-cluster family. However, before
extending further the discussion it is worth carrying out a similar
check for the M31 samples.

As noted in the previous section, in the list by
Huchra \etal (1991) the metal content only for a sub-sample of
the clusters lying in the selected area can be found. However, the weighted
mean
metallicity for the 68 \lxr globulars in M 31 turns out to be
($<[Fe/H]>_{\lxr}=-1.22\pm0.03$), in very good agreement with
the average values
obtained by Huchra \etal (1991) for all 150 clusters in their sample
($<[Fe/H]>=-1.21\pm0.02$). Hence, though far from being complete, the adopted
sample seems to be fairly representative of the whole set of clusters
observed in M 31 by Huchra \etal (1991).
\par\noindent
Based on the same data-set, the \lr clus\-ter po\-pu\-lat\-ion yields
a weigh\-ted mean\pn $<[Fe/H]>_{\lr}=-1.06\pm0.04$, {\it i.e.} higher
than the
\lxr group. The hypothesis that the two distributions are extracted from
the same parent population can be rejected at a $\sim 93\%$
($\sim 1.85 \sigma$) confidence level.

Admittedly, the statistical significance for the difference between the
two M31 cluster samples is lower than in the Galaxy. However, first,
it should be emphasized that both differences go in the same direction, and,
second, there are reasons to believe that the real difference could even be
larger. In fact, we must recall that: (i) since the surveyed area covers
mostly the inner regions of M31, the selected \lxr clusters could
represent the metal-rich tail of the clusters distribution in M31
(as shown for instance by the trend with the projected radius
displayed in their Figure 2 by Huchra \etal 1991), and (ii) the errors in
[Fe/H] are quite large in the M31 sample ($\sim \pm0.3~dex$), and
tend to smear out the difference.

In summary, though at a smaller level of significance, the
result found from the M31 samples is considered fully consistent with
that obtained for the Galaxy.
The comparison of the metallicity cumulative distributions
of the \lr and \lxr clusters are reported in  Figure 4$a,b$
for both the Galaxy (a) and M31 (b). \medskip\noindent

\medskip\noindent
{\bf 3.4. A two dimensional approach.}

\medskip\noindent
Since, as shown by Djorgovski (1991), in GGCs there are correlations between,
for instance, galactic position and metallicity
and central density and metallicity, two dimensional
KS-tests (Fasano and Franceschini, 1987; Press \etal, 1992) on the Galactic
data-sets were performed. The \lr and \lxr samples turn out to be
extracted from different parent population in any plane formed
by coupling the quoted parameters, at a confidence level always greater than
$ 99\%$. However, this result does not contribute any additional
information as it appears as a natural consequence of the combinations of
the differences separately detected via the one-dimensional tests and
the implicit correlations linking the driving parameters.

\bigskip\noindent
\centerline {\bf 4. DISCUSSION AND A NEW PROPOSED METALLICITY EFFECT}

\bigskip\noindent
The basic new result of the present paper for the samples of \lr and \lxr
clusters in the Galaxy and M31 is that in both galaxies the \lr clusters
seem to belong to a more metal-rich population than \lxr globulars.
The question is thus: Is metallicity a critical parameter to increase
the incidence of LMXBs or, since disk clusters are in general more metal-rich
than halo objects, the detected difference is just a by-product of
other primary effects?

\medskip\noindent
{\bf 4.1. Preliminary considerations}

\smallskip\noindent
So far there have been two basic answers
to the above question: (i) yes, the dependence on metallicity is just a
secondary by-product; (ii) no, it is germane, and (Grindlay 1987) it
is due to the claimed correlation between metallicity and the
slope of the cluster Initial Mass Function (IMF) (cf. McClure \etal
1986). Within this second scenario, metal-rich clusters would have
flatter IMF's and, in turn, more Neutron Stars (NS), the seeds to originate
LMXBs.

As it is well known, LMXBs are thought to be binary systems in which a NS
is accreting mass by the Roche lobe from a companion.
The first condition to fulfil to generate LMXBs is that a sufficient number
of NS's with velocities below the escape velocity from the cluster
are formed (Hut \etal 1991). Then, the NS should be somehow captured
to form a binary system. Till now, two main mechanisms have been
proposed (see for references Hut \etal 1991, 1992):
\smallskip\noindent
\item{\it (a)} 1+1 encounters ({\it i.e.} two-body tidal captures).
\par\noindent
\item{\it (b)} 2+1 encounters ({\it i.e.} exchange with a pre-existent binary
in which the lightest star is ejected from the system, leaving the NS bounded
in the new binary system).
\smallskip\noindent
The crucial parameter for both mechanisms seems to be the density of the
environment.

For example, with mechanism {\it (a)} a detailed calculation of the rate
of the basic phenomenon is available (see equation 3.6 in Lee and Ostriker,
1986), which shows that the rate of tidal capture increases by a factor
$\sim 10^6-10^7$ if stellar density in the clusters varies over the
range of densities covered by Galactic GCs. In particular, only for high
density clusters (with $Log \rho_0>4$) the mechanism is really efficient.
\noindent
Concerning the binary exchange mechanism, the situation is less
straightforward. Hut \etal (1991) report that mechanism {\it (b)} has  a
cross section which is more than two orders of magnitude larger than
that for NS capture by a main sequence star.
Davies, Benz and Hills (1993) agree and emphasize that,
if a GC contains a sufficiently large binary population, mechanism
{\it (b)} outnumbers the encounters expected via mechanism {\it (a)}.
Moreover, they present simulations of 2+1 encounters between a NS
and two different kind of binaries: tidal capture or hardened primordial
binaries. Their main conclusion is that the vast majority of NS-exchange
in binaries are generated by encounters involving primordial binaries.

On the other hand, for instance according to Bailyn (1993), the
cluster binary population is dominated by primordial systems
only in ``dynamically young''  clusters (sparse, low density GC)
while they were heavily destroyed (due to stellar encounters) in
``dynamically old'' systems (PCC or high density clusters).

This statement has been confirmed by Hut \etal (1992), who moreover point
out that, in order to compare the efficiency of the two mechanisms,
one has to take into account the rate of binary destruction, so that
``In PCC clusters most binaries will be ejected or
destroyed in less than an Hubble time. Collision products
(e.g. LMXBs, pulsars, CV's) are thus most likely formed by two body
processes''.

Adopting the framework summarized above, one may expect that LMXBs in high
density or PCC clusters are formed mainly via two body encounters
({\it i.e.} mechanism {\it a}). In the low-intermediate
density clusters, having still an high percentage of survived
primordial binaries, 2+1 encounters should be more efficient in
generating systems which can evolve into LMXBs. For instance, in the extreme
case of NGC 6712 (the lowest density cluster in the Galaxy containing a LMBX)
the LMXB observed could have a different origin with
respect to those found in high density clusters such as NGC 6441, NGC 7078,
etc.
In fact, NGC 6712 has a probability to form LMXBs via two body
encounters about $10^5-10^6$ times smaller than the high density clusters.

The idea that cluster star density plays a basic r\^ole in the
LMXB formation and evolution seems therefore beyond any reasonable doubt.
However, a new natural mechanism by which the detected
influence of metallicity on LMXB frequency could be explained as a ``germane''
important effect and not as a simple by-product is presented in what follows.

\medskip\noindent
{\bf 4.2. Testing possible metallicity effects}

\smallskip\noindent
Figure 5 shows the distribution of the Galactic \lr globulars
({\it filled dots}) with respect to the \lxr clusters
({\it open dots}) in a density-metallicity plane.
The plot appears peculiar: the cluster distribution has a well defined
triangular shape.
In fact, clusters with high metallicity and low density are clearly lacking.
This is probably due to a complex history of cluster formation, evolution
and survival. Though interesting, we postpone this discussion to a
forthcoming paper (Bellazzini \etal 1994, in preparation).

In order to get a quantitative indication of the influence of the
metal content on the LMXB formation rate in GCs we selected from the
available sample in Figure 5 the densest clusters
($Log \rho_0>5$). One obtains a roughly {\it isodense} sub-sample
where the two-body tidal capture is efficient and is probably the
dominant mechanism to produce LMXBs.
Then, this sub-sample was divided in two metallicity groups:
{\it the first}, including GCs with $[Fe/H]<-1$, and {\it the second}
with $[Fe/H]>-1$.

Finally, it is easy to compute the \lr to total globulars ratio
in the two metallicity groups ({\it i.e.} the relative frequency of \lr
clusters
in each metallicity bin):
$R_R={N_{\lr}}/{N_{tot}}=0.64 \pm 0.24$ for the metal-rich group and
$R_P={N_{\lr}}/{N_{tot}}=0.15 \pm 0.11$ for the metal-poor one.
The quoted errors have been calculated by propagating $1\sigma$-errors and
assuming a binomial distribution for $N_{\lr}$ and a Poisson distribution
for $N_{tot}$.
The further ratio $R_R/R_P$ between these two
ratios can eventually be considered to be a first rough estimate of the
impact (in terms of frequency of the LMXBs within an ``isodensity'' cluster
sample) due to increasing metallicity from $<[Fe/H]>\sim -1.67$ (metal-poor
group) to $<[Fe/H]>\sim -0.43$ (metal-rich group).
Using again propagation of errors, the best estimate $({R_R\over
R_P})_{best}=4.26 \pm 3.34$ is obtained.
Very schematically,
one gets that the formation of a LMXB in a metal-rich cluster would be
$\sim 4$ times more probable than in a metal-poor one. Nevertheless, because
of counting statistics on such a small sample, the permitted range of values
for $R_R/R_P$ turns out to span from a few hundreths to roughly 22,
with a reasonable $3\sigma$-upper limit of $\sim 15$. On the other hand, values
equal or less than 1 are ruled out by a simple $\chi^2$ test applied to
the sample of clusters with $Log \rho_0 >2$, in order to obtain a more robust
statistical significance.
The $H_0$ hypothesis that the probability of being \lr is the
same for the clusters more metal rich than $[Fe/H]=-1$ as for those
poorer than this limit is rejected with a confidence level greater than
$99.8 \%$, therefore substantially excluding that $R_R/R_P \le 1$.

This observational result ({\it i.e} $({R_R\over R_P})_{best}\sim 4$, with
$1<{R_R/R_P}\le 15$) can then be compared with theoretical predictions.

\medskip\noindent
{\it 4.2.1. Influence of the IMF}

\smallskip\noindent
Grindlay (1987,1993) suggested that the metallicity dependence of the
LMXB frequency comes via the possible dependence of the IMF slope on
[Fe/H] as claimed by McClure \etal (1986). In synthesis, metal-rich
clusters show a flatter mass function, are expected to produce more NS's  and,
as a consequence, they would have a larger rate of ``right'' binaries.

As stressed by Renzini and Fusi Pecci (1988) and
by Richer \etal (1991), due to both observational and theoretical problems,
the estimate of the ``true'' index of the IMF is particularly tricky and
uncertain. However, one must rely upon the available data.

To quantify the effect of the claim on the two-body capture rate,
we introduced the metallicity dependence of the IMF as proposed by McClure
\etal (1986) and computed the expected number of NS's computed according to
Hut \etal (1991). Following their assumptions, a 1.4 $M_{\odot}$ NS
is originated from a star with $M>8 M_{\odot}$.

Assuming for the two me\-tal\-li\-ci\-ty groups the mean me\-tal\-li\-ci\-ties
re\-por\-ted above,\pn
$N_{NS}^{M-R}/N_{NS}^{M-P}\sim 500$ is ob\-tained. Since the expected rate
of two body captures is directly proportional to the number of NS's
(Lee and Ostriker, 1986), one would compute that the probability of
having LMXBs in metal-rich clusters is $\sim 500$ times larger than
in metal-poor ones (of course, at fixed density). Such a ratio is hardly
compatible with the observed frequency even taking into account the many
uncertainties involved in the various steps.

In a recent analysis, Piotto (1993) and Djorgovski \etal (1993) found
a different dependence of the IMF on the cluster intrinsic parameters.
Adopting the monovariate relation (their eq. {\it 3b})
$${{\Delta x_0\over {\Delta [Fe/H]}}=-1.6}$$
where $x_0$ is the slope of the global IMF, one can easily repeat
the previous estimates getting $N_{NS}^{M-R}/N_{NS}^{M-P} \sim 500$ again, a
value still incompatible with observations.
\par
If one considers the global multivariate formula (their eq. {\it 5}), taking
into account also the claimed dependence of the IMF slope on $R_{GC}$ and
$|Z_{GP}|$, the result, leading to $N_{NS}^{M-R}/N_{NS}^{M-P} \sim 1000$,
is even more inconsistent with the observed LMXB frequencies.
Actually, the relation had to be extrapolated to a wider metallicity
range and this could be partially responsible for the discrepancy.
\par
However, a simple simulation shows that the assumption
$${{\Delta x_0} \over {\Delta [Fe/H]}} =-0.4$$ yelds
$N_{NS}^{M-R}/N_{NS}^{M-P} \sim 4$, in good agreement with the observed
$R_{R}/R_{P}$ ratio. Considering the quoted uncertainties, this slope
is constrained to be anyway $\geq -0.8$.
\par
In synthesis,
if this last metallicity-dependence for the IMF will be confirmed and if the
adopted formulas to compute the NS numbers are correct,
the suggestion put forward by Grindlay (1987, 1993) is thus a
plausible explanation for the increasing incidence of LMXBs with increasing
cluster metallicity. Moreover, one could also draw some other interesting
considerations.
\smallskip
First, turning the problem around, one could use the evidence that
the adoption of a steep dependence of the IMF slope on metallicity is
hardly compatible with the observed frequency of LMXB's in Galactic
globular clusters to conclude that such a dependence is much shallower
than the one proposed earlier (McClure \etal 1986).
\par
Second, since this flatter dependence by itself cannot account for the
observed IMF slope variation, this result, in turn, may support the whole
interpretative scenario proposed by Djorgovski \etal (1993) and Stiavelli
\etal (1992).
\par
In fact, these authors claim that the detected dependence of the IMF slope
--$x_0$-- on the galactic location can be accounted for considering
the effects of the interactions between clusters and Galactic
disk. In their view, all globular clusters are born with quite similar
IMF's and the detected differences in the Present Mass Functions (PMF)
are due to the subsequent dynamical evolution which actually steepens the
observed slope.
The number of NS's is presumably determined by the orginal IMF,
and not by the PMF, resulting after the dynamical interactions
affecting mostly the cluster low-mass members. The consistency
obtained between {\it (a)} the observed ratio of LMXBs with varying metal
content and {\it (b)} the number ratio predicted adopting a slope concerning
just the metallicity dependence equal to -0.4
could imply that this is the correct value to use to
describe the {\it intrinsic}  dependence of the IMF on metallicity.

\medskip\noindent
{\it 4.2.2. A ``natural'' effect}

\smallskip\noindent
Besides considering the effects induced by possible differences in the IMF's,
we suggest here that there is a {\it natural} factor
enhancing the probability of the two-body tidal capture
(and, in turn, the rate of LMXB
production) in high metallicity clusters. This {\it new} factor
comes from the stellar evolution theory itself.

In fact, the rate of tidal capture ($\Gamma$) as described by Lee and
Ostriker (1986) depends, at fixed cluster density, on radius ($R$) and mass
($M$) of the capturing star, and precisely:
$$\Gamma \sim R^{2-\alpha} M^{\alpha}~~~~~~~~~~~~~~~~(1)$$
where $\alpha =1.07$, a value appropriate for the case under consideration.

Following standard stellar models (VandenBerg and Bell, 1985),
stars with higher metal content have wider radii and higher masses
than metal-poor ones.
On the basis of those models, for a similar metallicity range,
and of the above Eq. 1, thus:
$${ {\Gamma_{M-Rich}} \over {\Gamma_{M-Poor}}} \sim 2.2$$
\noindent
This means that the {\it natural} contribution is at least
as important as that ot the IMF.

Furthermore, since for fixed binary separation and masses of the two
components, metal-rich stars would more easily fill the Roche lobe, as their
radii are larger, there is an additional ``evolutionary'' reason to favour
a higher incidence of LMXBs in metal-rich clusters.

\bigskip\noindent
\centerline {\bf 5. SUMMARY AND CONCLUSIONS}

\bigskip\noindent
Adopting appropriately selected samples of \lr and \lxr clusters in
the Galaxy and M31, it has been shown that:
\smallskip\noindent
\item{1.}  The \lr clusters (with $L_X > 10^{36} erg/sec$),
commonly believed to contain the so-called Low Mass X-ray Binaries, are,
in both galaxies, in
general denser than \lxr globulars, as expected on the basis of
current LMXB models. Furthermore,  both \lr cluster
samples in the Galaxy and M31 are
shown to be significantly metal richer than the corresponding \lxr ones.
\par\noindent

\item{2.} Within this framework and based on the accurate testing  of the
multivariate formula of Djorgovski \etal (1993), (eq. $5$),
the apparent larger frequency
of X-ray clusters at low $R_{GC}$ and $|Z_{GP}|$ is simply a  by-product
of the primary dependencies on density and metallicity.

\smallskip\noindent

In order to test possible mechanisms through which metallicity can act as a
critical parameter in increasing the LMXBs frequency at a fixed cluster
density,
has been adopted a framework where LMXBs in
GCs are formed via two main mechanisms:
(i) two body tidal captures form bound systems including a Neutron Star,
in high density GCs, heavily depleted of primordial binaries, and (ii)
formation of binaries via (2+1) encounters in low-density clusters, where
primordial binaries are presumably still present.

The larger observed incidence of LMXBs in metal-rich
clusters can be explained by the predicted increase of available Neutron Stars
(Grindlay, 1987,1993), but the slope of the metallicty dependence of the IMF
is constrained to be ${{\Delta x_0}\over {\Delta [Fe/H]}} > -0.8$ to preserve
compatibility with data.

Second, there is also a {\it natural} factor
enhancing the probability of the two-body tidal capture (and, in turn,
the rate of LMXB production) in high metallicity clusters which it is
suggested to come from the stellar evolution theory itself.

{}From the rough computations made, this new effect could by
itself explain the observed ratio of 4 between the incidences of \lr clusters
in the metal-rich and the metal-poor groups determined above. However,
there is no reason to exclude that both mechanisms proposed so far to explain
the dependence of the LMXB frequency on metallicity can be at work.

Finally, it may be interesting to note that the metallicity-dependence of
the IMF slope, if real, would be efficient
in enhancing the frequency of LMXBs independent of the actual
mechanism adopted to form the LMXBs as it would directly increase
the number of available NS's. On the contrary, the ``natural''
enhancement suggested here is {\it framework-dependent}
as it would be most efficient in the two-body tidal captures.
On the other hand, unless the evolutionary models are grossily in error,
it must be surely present and the rate of captures should always
increase with increasing metallicity of the star interacting with
the NS.

In conclusion, the present results and suggestions represent
an additional item in a scenario (see f.i. Fusi Pecci \etal 1992, 1993c,d,
Djorgovski and Piotto 1992, Ferraro \etal 1993) where global cluster
dynamical effects and evolutionary properties of individual stars
interfere to yield special objects or features which may be typical of
very crowded systems.

\bigskip\noindent
\bigskip\noindent
\bigskip\noindent
{\sl Acknowledgements.}~~
FFP warmly thanks Josh Grindlay and Alvio Renzini for many useful
discussions during which the essence of the problem has been
analysed and clarified. We are also grateful to Giorgio Palumbo for a critical
reading of the manuscript. The financial support of the {\it Agenzia Spaziale
Italiana} is kindly acknowledged.

\bigskip
\bigskip

\parskip=0pt

\centerline{\bf References}
\bigskip\noindent
\pp Armandroff, T.E., 1989. AJ, 97, 375

\pp Armandroff, T.E., \& Zinn, R.J. 1988, AJ, 96, 92

\pp Armandroff, T.E., Da Costa, G., \& Zinn, R.J. 1992, AJ, 104, 164

\pp Bailyn, C.D. 1993, Structure and Dynamics of Globular Clusters,
ed. S.G. Djorgovski \& G. Meylan, ASP Conf. Ser., 50, 191

\pp Battistini, P., B\`onoli, F., Braccesi, A., Fusi Pecci, F., \& Marano, B.
1980, A\&AS, 42, 357

\pp Battistini, P., B\`onoli, F., Buonanno, R., Corsi, C.E., \& Fusi Pecci, F.
1982, A\&A, 113, 39

\pp Battistini, P., B\`onoli, F., Braccesi, A., Federici, L., Fusi Pecci, F.,
Marano, B., \& B\"orngen 1987, A\&AS, 67, 447

\pp Battistini, P., B\`onoli, F., Casavecchia, M., Ciotti, L., Federici, L.,
\& Fusi Pecci, F. 1993, A\&A, 272, 77

\pp Battistini, P. \etal 1994, (in preparation)

\pp Buonanno, R., Corsi, C.E., Battistini, P., Bonoli, F., \&
Fusi Pecci, F. 1982, A\&AS, 47, 451

\pp Cohen, J.G., \& Freeman, K.C. 1991, AJ, 101, 483

\pp Crampton, D., Cowley, A.P., Hutchings, J.B., Schade, D.J., \&
van Speybroeck, L. 1984, ApJ, 284, 663 (CCHSV)

\pp Crampton, D., Cowley, A.P., Schade, D.J., \& Chayer, P. 1985,
ApJ, 288, 494

\pp Davies, M.B., Benz, W., \& Hills, J.G. 1993, Structure and Dynamics
of Globular Clusters, ed. S.G. Djorgovski \& G. Meylan, ASP Conf. Ser.,
50, 185

\pp Djorgovski, S.G. 1991, Formation and Evolution of Star Clusters, ASP Conf.
Ser., 13,112

\pp Djorgovski, S.G. 1993, Structure and Dynamics of Globular Clusters,
ed. S.G. Djorgovski \& G. Meylan, ASP Conf. Ser., 50, 373

\pp  Djorgovski, S.G., \& Piotto, G. 1992, AJ, 104, 2112

\pp Djorgovski, S.G., Piotto, G., \& Capaccioli, M. 1993, AJ, 105, 2148

\pp Fasano, G., \& Franceschini, A. 1987, MNRAS, 225, 155

\pp Federici, L., B\`onoli, F., Ciotti, L., Fusi Pecci, F., Marano, B.,
Lipovetsky, V.A., Neizvestny, S.L., \& Spassova, N. 1993, A\&A, 274, 87

\pp Ferraro, F.R., Fusi Pecci, F., Cacciari, C., Corsi, C.E., Buonanno, R.,
Fahlman, G.G., \& Richer, H.B., 1993, AJ, 106, 2324

\pp Fusi Pecci, F., Ferraro, F.R., Corsi, C.E., Cacciari, C., \&
Buonanno, R. 1992, AJ, 104, 1831

\pp Fusi Pecci F., Battistini, P., Bendinelli, O., B\`onoli, F.,
Cacciari, C., Djorgovski, S.G., Federici, L., Ferraro, F.R., Parmeggiani,
G., Weir, N., \& Zavatti, F. 1993a, A\&A, 284, 349

\pp Fusi Pecci, F., Cacciari, C., Federici, L., \& Pasquali, A. 1993b,
The Globular Cluster-Galaxy Connection,
ed. G.H. Smith \& J.P. Brodie, ASP Conf. Ser., 48, 410

\pp Fusi Pecci, F., Ferraro, F.R., Bellazzini, M. Djorgovski, S.G.,
Piotto, G., \& Buonanno, R. 1993c, AJ, 105, 1145

\pp Fusi Pecci, F., Ferraro, F.R., \& Cacciari, C., 1993d, Blue Stragglers,
ed. M. Livio and R.  Saffer, ASP Conf. Ser., 53, 97

\pp Grindlay, J.E. 1987, Origin and Evolution of Neutron Stars,
IAU Symp. No. 125, ed. D. Helfand \& J. Huang, Dordrecht: Reidel,  p.173

\pp Grindlay, J.E. 1993, The Globular Cluster-Galaxy Connection,
ed. G.H. Smith \& J.P. Brodie, ASP Conf. Ser., 48, 156

\pp Huchra, J.P., Brodie, J.P., \& Kent, S.M.  1991, ApJ, 370, 495

\pp Hut, P., Murphy, B.W., \& Verbunt, F. 1991, A\&A, 241, 137

\pp Hut, P., McMillan, S., Goodman, J., Mateo, M., Phinney, E.S., Pryor,C.,
Richer, H.B., Verbunt, F., \&  Weinberg,M. 1992, PASP, 104, 981

\pp Lee, H.M., \& Ostriker, J.H. 1986, ApJ, 310, 176

\pp Lightman, A., \& Grindlay, J.E. 1982, ApJL, 262, 145

\pp Long, K.S., \& van Speybroeck, L.P. 1983, Accretion Driven Stellar X-ray
Sources,  ed. Lewin, W. \& E.P.J. van den Heuvel, Cambridge University Press,
p. 41 (LVS)

\pp McClure, R.D., VandenBerg, D.A., Smith, G.H., Fahlman, G.G., Richer, H.B.,
Hesser, J.E., Harris, W.E., Stetson, P.B., \& Bell, R.A. 1986, ApJL, 307, L49

\pp Piotto, G. 1993, Structure and Dynamics of
Globular Clusters, ed. S.G. Djorgovski \& G. Meylan, ASP Conf. Ser.,
50, 233

\pp Press, W.H., Teukolsky, S.A., Vetterling, W.T., Flannery, B.P., {\it Numeri
cal Recipes in FORTRAN}, 2nd Ed., Cambridge Univ. Press, 1992 (Sec. 14.7)

\pp Primini, F.A., Forman, W., \& Jones, C. 1993, ApJ, 410, 615 (PFJ)

\pp Pryor, C., \& Meylan, G. 1993, Structure and Dynamics of
Globular Clusters, ed. S.G. Djorgovski \& G. Meylan, ASP Conf. Ser.,
50, 357

\pp Renzini, A., \& Fusi Pecci, F. 1988, ARA\&A, 26, 199

\pp Richer, H.B., Fahlman, G.G., Buonanno, R., Fusi Pecci, F, Searle, L.,
\& Thompson, I.B. 1991, ApJ, 381, 147

\pp Stiavelli, M., Piotto, G.P., \& Capaccioli, M., 1992, in Morphological
and Physical Classification of Galaxies, edited by G. Longo, M. Capaccioli, and
G. Busarello, (Kluwer, Dordrecht), p. 455

\pp Trinchieri, G., \& Fabbiano, G. 1991, ApJ, 382, 82 (TF)

\pp VandenBerg, D.A., \& Bell, R.A. 1985, ApJS, 58, 561

\pp Zinn, R.J. 1985, ApJ, 293, 424.

\vfill
\eject

\centerline{\bf Figure Captions:}

\bigskip
\noindent{\bf Figure 1:}~Log-central density cumulative distributions
($\Phi$) of the \lxr ({\it dashed line}) and \lr ({\it solid line}) globular
clusters in the Galaxy ({\it panel a}) and in M31 ({\it panel b}),
respectively.
Note that in ({\it panel a}) $\rho_{0}$ is the adopted cluster central
density, while in ({\it panel b}) W$_{1/4}$ is the half-width at $1/4$ of the
height of the two-dimensional fit of the cluster image (normalized to
an arbitrary scale) which is at first order anti-correlated to the
cluster central density (see Sections 2 and 3.1).

\bigskip
\noindent{\bf Figure 2:}~Radial cumulative distributions ($\Phi$) of the
\lxr ({\it dashed line}) and \lr ({\it solid line}) globular
clusters in the Galaxy ({\it panel a}) and in M31 ({\it panel b}),
respectively (units in Kpc).
Note that in {\it panel a} R$_{GC}$ is the adopted
distance from the Galactic center, while in {\it panel b} R$_{M31projected}$
is the projected distance (see Sections 2 and 3.2).

\bigskip
\noindent{\bf Figure 3:} Cumulative distributions ($\Phi$)
of the \lxr ({\it dashed line}) and \lr ({\it solid line}) globular clusters in
the Galaxy with respect to ~$|Z_{GP}|$ (Kpc) --the height on the Galactic
Plane.
A zoomed plot has also been inserted to show that most of the difference
beetween the two distributions is actually found for very small values
of ~$|Z_{Gp}|$. In particular, $75\%$ of the \lr sample is located
within the first Kpc from the Galactic Plane.

\bigskip
\noindent{\bf Figure 4:}~Metallicity cumulative distributions
($\Phi$) of the \lxr ({\it dashed line}) and \lr ({\it solid line}) globular
clusters in the Galaxy ({\it panel a}) and in M31 ({\it panel b}), respectively
(see Sections 2 and 3.3).

\bigskip
\noindent{\bf Figure 5:}~The distribution of the X-ray clusters ({\it filled
dots}) and the non-X-ray globulars ({\it open dots}) in the Galaxy in
a Log-density -- metallicity plane.

\end
\bye